# The 2003 Nov 14 occultation by Titan of TYC 1343-1865-1. II. Analysis of light curves

A. Zalucha[1], A. Fitzsimmons[2], J. L. Elliot[1,3], J. Thomas-Osip[4], H. B. Hammel[5], V. S. Dhillon[6], T. R. Marsh[7], F. W. Taylor[8], P. G. J. Irwin[8]

[1]Dept. of Earth, Atmospheric, and Planetary Sciences, Massachusetts Institute of Technology, 77 Massachusetts Ave., Room 54-1721, Cambridge, MA 02139, USA
[2]Dept. of Physics and Astronomy, Queen's University Belfast BT7 1NN, Northern Ireland
[3]Dept. of Physics, Massachusetts Institute of Technology, 77 Massachusetts Ave., Room 54-422, Cambridge, MA 02139, USA; and Lowell Observatory, 1400 West Mars Hill Road, Flagstaff, AZ 86001, USA
[4]Las Campanas Observatory, Observatories of the Carnegie Institute of Washington, Casilla 601, La Serena, Chile
[5]Space Science Institute, 4750 Walnut Street, Suite 205, Boulder, CO 80301, USA
[6]Dept. of Physics and Astronomy, University of Sheffield, Sheffield S3 7RH, UK
[7]Dept. of Physics, University of Warwick, Coventry CV4 7AL, UK
[8]Atmospheric, Oceanic, and Planetary Physics, University of Oxford, Clarendon Laboratory, Parks Rd, Oxford, OX1 3PU, UK

Number of manuscript pages: 58
Number of figures: 14
Number of Tables: 5
Running head: Analysis of the 2003 Nov 14 occultation by Titan

Name and address to which editorial correspondence and proofs should be directed:
Angela Zalucha, MIT
77 Massachusetts Ave., Room 54-1721
Cambridge, MA  02139
azalucha@mit.edu

We observed a stellar occultation by Titan on 2003 November 14 from La Palma Observatory using ULTRACAM with three Sloan filters: $u'$, $g'$, and $i'$ (358, 487, and 758 nm, respectively). The occultation probed latitudes 2°S and 1°N during immersion and emersion, respectively. A prominent central flash was present in only the $i'$ filter, indicating wavelength-dependent atmospheric extinction. We inverted the light curves to obtain six lower-limit temperature profiles between 335 and 485 km (0.04 and 0.003 mb) altitude. The $i'$ profiles agreed with the temperature measured by the Huygens Atmospheric Structure Instrument [Fulchignoni, M., and 43 colleagues, 2005. Nature 438, 785–791] above 415 km (0.01 mb). The profiles obtained from different wavelength filters systematically diverge as altitude decreases, which implies significant extinction in the light curves. Applying an extinction model [Elliot, J.L., Young, L.A., 1992. Astron. J. 103, 991–1015] gave the altitudes of line of sight optical depth equal to unity: $396 \pm 7$ km and $401 \pm 20$ km ($u'$ immersion and emersion); $354 \pm 7$ km and $387 \pm 7$ km ($g'$ immersion and emersion); and $336 \pm 5$ km and $318 \pm 4$ km ($i'$ immersion and emersion). Further analysis showed that the optical depth follows a power law in wavelength with index $1.3 \pm 0.2$. We present a new method for determining temperature from scintillation spikes in the occulting body's atmosphere. Temperatures derived with this method are equal to or warmer than those measured by the Huygens Atmospheric Structure Instrument. Using the highly structured, three-peaked central flash, we confirmed the shape of Titan's middle atmosphere using a model originally derived for a previous Titan occultation [Hubbard, W.B., and 45 colleagues, 1993. Astron. Astrophys. 269, 541–563].

## 1. Introduction

The stratospheric and mesospheric region of Titan's atmosphere (roughly 200 to 600 km altitude) is perhaps its least-well studied portion. Recent *in situ* data from the Huygens Atmospheric Structure Instrument (HASI) ranged from 0 to 1400 km (Fulchignoni et al., 2005). Two sets of temperature data from the Composite Infrared Spectrometer (CIRS) instrument onboard Cassini ranged from 100 km to about 500 km (Vinatier et al., 2007); data from the Cassini Ultraviolet Imaging Spectrometer (UVIS) ranged from 450 to 1600 km (Shemansky et al., 2005). Griffith et al. (2005) developed a simple model of the temperature in the region between 300 and 600 km using $CH_4$ spectra. The 28 Sgr occultation by Titan in 1989 (Sicardy et al., 1990; Hubbard et al., 1990) yielded temperature profiles for the 300 to 500 km region, the (non-spherical) shape of the stratosphere, measurements of the stratospheric zonal winds, and measurements of haze properties (Hubbard et al., 1993). Tracadas et al. (2001) derived an isothermal atmospheric model from an occultation that occurred in 1995, and Bouchez et al. (2003) inferred a non-spherical shape for the mid-latitude stratosphere from an occultation that occurred in 2001.

Occultation data continue to be an important component to our understanding of the properties (e.g., temperature, density, pressure) of Titan's middle atmosphere. Here we report the analysis of simultaneous, three-color, high time resolution light curves of the 2003 November 14 occultation of Tycho 1343-1865-1 (Fitzsimmons et al., 2007). The high time resolution of our observations allowed for the analysis of sharp spikes in the light curves caused by density features in Titan's atmosphere. The proximity of the occultation track to the center of Titan's disk permitted the detection of a central flash, from which the shape of Titan's atmosphere could be determined. Simultaneous, multi-wavelength observations at the same location allowed for the study of the dependence of atmospheric extinction on wavelength and altitude.

## 2. Data

Fitzsimmons et al. (2007; hereinafter referred to as Paper I) presented the measurements and calibration procedure in more detail; we include a summary of the key points here. Titan occulted the star Tycho 1343-1865-1 on 2003 November 14 from 06:49:11 UT to 06:58:39 UT. Figure 1 shows the star's path around Titan's limb. We observed the event from the island of La Palma, Spain using the ULTRACAM instrument (Dhillon et al., 2007). ULTRACAM provided very high time resolution at 30 points per second for this dataset. Simultaneous measurements were available at three wavelengths: Sloan filters *u'*, *g'*, and *i'*, which—after including the effects of the terrestrial atmosphere—were centered at effective wavelengths 358, 487, and 758 nm, respectively (Paper I). The overlap between the *u'* and *g'* filters of ULTRACAM was about 20 nm; there was no overlap between *g'* and *i'*. In the calibrated light curves, about 240 spikes were present. We detected a highly structured, three-peaked central flash in the *i'* filter, but not in the other two.

*<u>*Insert Fig. 1</u>*

## 3. Light Curve Model Fitting

We used the light curve model of Elliot and Young (1992; hereinafter EY92) to model our data. This light curve model assumes a temperature structure in the form of

$$T(r) = T_H \left( \frac{r}{r_H} \right)^b , \qquad (1)$$

where $T$ is temperature, $T_H$ is the half-light temperature, $r$ is the radius from the body's center, $r_H$ is the half-light radius, and $b$ is the thermal gradient parameter. This particular structure was chosen because it gives an analytical expression for the light curve. Another parameter that describes the atmospheric structure is the lambda parameter $\lambda$, which is the ratio of gravitational potential energy to molecular thermal energy and is given by

$$\lambda(r) = \frac{\mu m_{amu} G M_p}{r k T(r)}, \tag{2}$$

where $\mu$ is the mean molecular weight, $m_{amu}$ is the atomic mass unit, $G$ is the gravitational constant, $M_p$ is the mass of the occulting body, and $k$ is the Boltzmann constant.

The EY92 model assumes an exponential dependence of the linear absorption coefficient, $\kappa$, on the distance from Titan's center of the form

$$\kappa(r) = \kappa_1 \exp\left(-\frac{r - r_1}{H_{\tau 1}(r/r_1)}\right) \qquad (r \leq r_1) \tag{3}$$

$$\kappa(r) = 0 \qquad (r > r_1),$$

where $\kappa_1 = \kappa(r_1)$. The parameters $r_1$, $r_2$, and $H_{\tau 1}$ are the altitude of extinction onset (an abrupt haze turn-on radius), the radius of optical depth unity, and the haze scale height at the radius of extinction onset, respectively. As discussed by EY92, the line of sight optical depth, $\tau(r)$, is:

$$\tau(r) = \left(\frac{r}{r_2}\right)^{\frac{3}{2}} \exp\left[\frac{r_1^2}{H_{\tau 1}}\left(\frac{1}{r} - \frac{1}{r_2}\right)\right] \frac{\mathrm{erf}\left[\frac{r_1}{r}\left(\frac{r_1^2 - r^2}{2 H_{\tau 1} r}\right)^{\frac{1}{2}}\right] C(\delta_\tau)}{\mathrm{erf}\left[\frac{r_1}{r_2}\left(\frac{r_1^2 - r_2^2}{2 H_{\tau 1} r_2}\right)^{\frac{1}{2}}\right] C(\delta_{\tau 2})}; \qquad (r \leq r_1) \tag{4}$$

$$\tau(r) = 0 \qquad (r > r_1).$$

The parameters $\delta_\tau$ and $\delta_{\tau 2}$ are defined by

$$\delta_\tau = H_{\tau 1} r / r_1^2, \tag{5}$$

$$\delta_{\tau 2} = H_{\tau 1} r_2 / r_1^2 , \tag{6}$$

and $C(\delta_\tau)$ is a power series of which the first five terms are:

$$C(\delta_\tau) = 1 + \frac{9}{8} \delta_\tau + \frac{345}{128} \delta_\tau^2 + \frac{9555}{1024} \delta_\tau^3 + \frac{1371195}{32768} \delta_\tau^4 + \dots \tag{7}$$

(likewise for $C(\delta_{\tau 2})$).

The light curves used in the model fitting were those obtained following the data acquisition and calibration procedures described in Paper I. The times under consideration for this analysis were between 2600.968 and 3797.573 s after 06:00:00 UT (hereinafter all times are given relative to this zero point). Our first goal was to model the overall structure of the light curves; thus, they were first averaged over 60 points (2 s) as a smoothing operation, yielding a 598 point fit for each filter.

We fit the light curves by least squares using the EY92 model. For this analysis, $b$ was held fixed to prevent the other model parameters from reaching physically implausible values. A value of –1.5 was chosen based on examining the HASI temperature profile (Fulchignoni et al., 2005). In addition to the atmospheric parameters described above (i.e., $r_H$, $b$, $r_1$, $r_2$, and $H_{\tau 1}$), model parameters include the distance of closest approach of the telescope to the center of Titan's shadow on Earth $y_{min}$, time of closest approach (or occultation mid time) $t_{mid}$, and half-light equivalent isothermal lambda $\lambda_{Hi}$. This parameter is defined as the lambda parameter (Eq. 2) of an isothermal atmosphere that would produce a light curve with the same slope at half-light as the nonisothermal atmosphere. Mathematically, the half-light equivalent isothermal lambda corresponds to

$$\lambda_{Hi} = \frac{\mu m_{amu} G M_p}{r_H k T_H} + \frac{5b}{2} . \tag{8}$$

The background parameters of light curve model fitting differ somewhat from those of EY92. We assume the observed signal as a function of time, $s(t)$, is given by

$$s(t) = [\zeta(t)(1-s_{bf}) + s_{bf}][s_f + s'(t-t_{mid})], \tag{9}$$

where $\zeta(t)$ is the normalized light curve (which depends on atmospheric parameters and occultation geometry), $s_{bf}$ is the background fraction, $s_f$ is the full-scale level, $s'$ is the background slope, and $t_{mid}$ is the occultation mid time. By writing the equation for $s(t)$ in this way, we assume that all the drifts in $s(t)$ are due to linear changes in the extinction of starlight and Titan light by Earth's atmosphere, and that the main component of the background is the light from Titan. In order to ensure $s(t) = \zeta(t)$ when there is no background (i.e., $s_{bf}$ and $s'$ are zero), we set $s_f = 1$. Table 1 lists the fit parameters for each light curve; Fig. 2 displays the model curves with the data used to obtain the model. Figure 3 shows the light curves after correcting for the background.

*Insert Table 1*

*Insert Fig. 2*

*Insert Fig. 3*

In Table 1, the radii of extinction onset are all greater than the half-light radii. The time of closest approach agrees in all three filters, as it should since this parameter is not dependent on wavelength. The error bars on the $u'$ values are generally larger since that light curve has more scatter. The $i'$ errors are larger than the $g'$ since an additional parameter (the distance of closest approach) was fit in the $i'$ light curve because it has a well defined central flash. For the atmospheric extinction parameters, the altitude of extinction onset and of optical depth unity increase from the $i'$ to the $u'$ filters. This behavior supports the conclusion that the extinction increases with decreasing wavelength. The values obtained here for the extinction parameters should be used with caution; they will be discussed further in §5.

The half-light radii shown in Table 1 are about 150–200 km higher than the optical limb of Titan, which is around 200–250 km above the surface. Also, our half-light radii—i.e., $2992 \pm 13$ km found for the *i'* filter—are consistently smaller than the half-light radii derived from the occultation of 28 Sgr: $3031 \pm 10$ km as determined by Sicardy et al. (1990) and $3031 \pm 6$ km as determined by Hubbard et al. (1990). One possible reason for the discrepancy is that Sicardy et al. (1990) and Hubbard et al. (1990) assumed an isothermal, extinction-free atmosphere to determine their values whereas we did not. Our model defines the half-light radius as that which would be observed in the absence of extinction, so a direct comparison between our value and that derived for the 28 Sgr occultation is possible. Isothermal fits to the light curves were very poor and yielded physically implausible temperatures ranging from 19–27 K. An isothermal fit with extinction yielded slightly higher half-light radii than the nonisothermal fit of Table 1, but not enough to account for the discrepancy. The effect of extinction may be understood qualitatively as follows: an atmosphere with extinction would produce a light curve that drops off faster than a clear atmosphere. The half-light time (i.e., when the normalized flux is 0.5) then occurs sooner during immersion and later during emersion. These times correspond to higher altitudes in the atmosphere, since altitude (and radial distance from the body's center) increases as the observer moves farther away in time from the mid time of the occultation. Failing to account for extinction thus causes higher half-light radii to be derived, thus explaining why Sicardy et al. (1990) and Hubbard et al. (1990) could have obtained higher half-light radii.

## 4. Light Curve Inversion

We performed a total of six light curve inversions—three for the immersion portions and three for the emersion portions of the background-corrected light curves—using the method of Elliot et al. (2003). The upper boundary condition was determined by least squares fitting the portion of the light curve above the half-light level to a similar model as described in §3 (but with no extinction parameters). We then converted the remainder of the light curve from flux as a function of time to refractivity as a function of altitude. This method assumes the refractivity is proportional to the number density, from which pressure and temperature may be obtained from hydrostatic balance and the ideal gas law. The method also assumes that the atmosphere is clear; i.e., that the light curve is free of extinction effects.

Table 2 lists the light curve inversion parameters. The shadow velocity, distance of closest approach, and time of closest approach that were also required for the inversion were the same as the $i'$ values in Table 1. The Gas Chromatograph Mass Spectrometer (GCMS) onboard the Huygens spacecraft confirmed that Titan's atmosphere is primarily composed of $N_2$ (Niemann et al., 2005); thus we assumed the atmosphere was entirely $N_2$. We took the error in the flux for each inversion to be the standard deviation of the 2000 baseline points with the highest shadow plane distance from Titan's center (each immersion half contained roughly 19000 points and each emersion half contained roughly 17000 points before averaging). We binned the un-averaged, background-corrected data (with the same time span as in the light curve modeling procedure) by 10 points (0.33 s) in the time domain and then averaged it such that each point had a positive flux. We performed the latter type of averaging because the inversion method required positive flux. When a negative flux point was encountered, it was averaged with successive points until a positive flux was achieved (see §3 of Elliot et al., 2003). Table 3 lists the number of points after averaging in the boundary region and in the inversion region for each inversion calculation.

Figure 4 shows the resulting limiting temperature profiles and their temperature error for each filter. The errors are the formal errors that arise from the least squares fitting procedure used to obtain the boundary conditions. The arrows in Fig. 4 indicate that the profiles are a lower limit on temperature, since extinction effects in the light curve, if present, would cause the inversion procedure to give a lower temperature than the actual temperature of the atmosphere at a given altitude.

Figure 5 re-displays all six limiting temperature profiles with the error bars omitted for clarity, along with the temperature profile from HASI (Fulchignoni et al., 2005). The inverted curves have the same overall shape, but are displaced from the others in temperature as a function of wavelength. In a clear atmosphere, the immersion profiles should match each other within their error bars, as should the emersion profiles, since atmospheric temperature is independent of the filter wavelength. Elliot et al. (2003) demonstrated that an incorrect calibration of the light curve (i.e., a systematic error in the background level), can result in a false temperature gradient. We assume any calibration errors that may exist are not related in such a way as to create a wavelength dependence in the inversion profiles.

Atmospheric extinction would cause inversion temperatures to be displaced to colder temperatures than the actual temperature. Furthermore, wavelength-dependent extinction would make the inversion temperatures diverge. If the warmest profile, the $i'$ profile, has the least extinction and is thus closest to the true atmospheric temperature, then the $g'$ and emersion $u'$ curves must contain the most extinction effects since they are displaced from the $i'$ curve to colder temperatures (for a constant altitude). The $i'$ profile agrees with the HASI temperatures above about 415 km, implying that the extinction at this wavelength is negligible above this altitude. At lower altitudes the $i'$ profile also becomes colder than the HASI temperatures, which indicates significant extinction.

The presence of extinction is further supported by examining Fig. 2, where the central flash is clearly visible in only the $i'$ curve. Since the refractivities at the wavelengths under consideration are nearly the same, the central flash should be identical in each light curve. Presumably, the flux from the central flash at shorter wavelengths is weakened by extinction to the point of being undetectable.

Each immersion and emersion profile pair (for a given filter) in Fig. 5 should match since they probe Titan at nearly the same latitude and longitudinal variations are assumed negligible (Fig. 1). An apparent disagreement in temperature of about 30 K exists between the $i'$ profiles below 380 km altitude; however this disagreement may be a manifestation of a disagreement in the altitude scales. The formal error in the altitude (not shown in Figs. 4 or 5) resulting from errors in the boundary condition is less than 2 km, which does not account for the 10–15 km altitude difference (at constant temperature) in the inversion profiles. Elliot et al. (2003) showed that a systematic error in the background flux as small as 1% could cause an error in the altitude scale of up to one third of a scale height, which is about 15–20 km for Titan's atmosphere. However, the main result of Fig. 5—that extinction increases with decreasing wavelength—remains valid.

## 5. Extinction Modeling

Figure 6 shows a diagram of the coordinate system used in the following calculations. If the refraction angle, $\theta$, as a function of the distance from the occulting body's center, $r$, is known, the extinction-free light curve can be generated via

$$\phi(r) = \left(\frac{1}{1 + D[d\theta(r)/dr]}\right)\left(\frac{1}{1 + D[\theta(r)/r]}\right), \tag{10}$$

where $D$ is the distance from the observer to the occulting body and flux $\phi(r)$ from perpendicular limb points should be summed (see Elliot and Olkin, 1996 for more details). The light curve with extinction is then given by

$$\zeta(r) = \phi(r)e^{-\tau_{obs}(r)}, \tag{11}$$

where $\tau_{obs}(r)$ is the optical depth along the path of the ray. By dividing the measured light curve, $\zeta(r)$, by $\phi(r)$, one can extract $\tau_{obs}(r)$.

*Insert Fig. 6*

In the present case, we calculated the extinction-free light curve, $\phi(r)$, from HASI data (Fuchignoni et al., 2005), shown in Fig. 7. The Huygens probe gathered *in situ* data as it descended through Titan's atmosphere 14 months after our occultation occurred. The probe landed at approximately 10°S, compared to our nearly equatorial observations. The combination of these factors makes it the best dataset to compare with the occultation data. Because the occultation did not probe below 250 km, we considered the portion of the HASI profile that extends from 157 km to 1380 km altitude, which was calculated from density measurements as the Huygens spacecraft decelerated. We fit the amplitudes of a Fourier series with 40 frequencies using the method of least squares to the HASI temperature data to get a smooth, analytical temperature curve as a function of $r$. The logarithm of the corresponding pressure data was interpolated using a cubic polynomial to derive pressure as a function of radius. From these temperature and pressure functions, the number density, $n(r)$, may be found from the ideal gas law. The refractivity can be derived from

$$\nu(r) = \frac{\nu_{STP} n(r)}{L}, \qquad (12)$$

where $\nu_{STP}$ is the refractivity of $N_2$ at standard temperature and pressure (given in Table 2) and $L$ is Loschmidt's number. Then $\theta(r)$ and $d\theta(r)/dr$ may be determined from Eqs. 2 and 3 of Chamberlain and Elliot (1997):

$$\theta(r) = \int_{-\infty}^{\infty} \frac{r}{r'} \frac{d\nu(r')}{dr} dx \qquad (13)$$

and

$$\frac{d\theta(r)}{dr} = \int_{-\infty}^{\infty} \{\frac{x^2}{(r')^3} \frac{d\nu(r')}{dr'} + \frac{r^2}{(r')^2} \frac{d^2\nu(r')}{dr'^2}\} dx \qquad (14)$$

where $r'$ is an integration constant and $x$ is the distance along the ray path. By examining Fig. 6, it can be seen that $r'^2 = r^2 + x^2$. The extinction-free light curves (one for each wavelength) can now be derived from Eqs. 10 and 12–14 and are shown in Fig. 8. Also shown in the figure for comparison are the model fits to the observed light curves from Fig. 3. To display the light curves as a function of time, $t$, the following relations in terms of shadow plane distance $y$ were used:

$$y = r + D\theta(r) \qquad (15)$$

and

$$y = \sqrt{y_{min}^2 + V^2(t - t_{mid})^2}, \qquad (16)$$

where $y_{min}$ is the distance of closest approach in the shadow plane, $V$ is the relative velocity between the body and the observer, and $t_{mid}$ is the time of closest approach.

The measured (background-corrected) light curve data were averaged over a 2-second period (i.e., the same as in §3 and the points in Fig. 3). These were taken to be $\zeta(r)$. The extinction term $e^{-\tau_{obs}(r)}$ was then calculated from Eq. 11. Because of scatter in the data due to noise, this term had negative values, which leads to imaginary values for $\tau_{obs}(r)$. As in the inversion procedure, a negative point must be successively averaged with nearby points until a positive value is reached. After this step we then computed $\tau_{obs}(r)$, which is the optical depth required to make the measured light curves and hence the light curve inversion temperatures agree with the HASI temperature profile (Fig. 5).

Next we examined $\tau_{obs}(r)$ using the extinction model of EY92, which specifies the optical depth as Eq. 4. This model was originally developed for Pluto's atmosphere, but we use it here to take advantage of the already existing model framework. In the light curve modeling procedure discussed in §3, which uses this extinction model, the haze parameters are coupled with other light curve parameters, especially the half-light radius and lambda parameter. In this section of the analysis, we consider the haze parameters after they have been separated from the light curve.

We least squares fit $\tau_{obs}(r)$ to Eq. 4 with a single free parameter $r_2$, which implicitly depends on wavelength. Again we performed a total of six individual fits corresponding to the immersion and emersion light curve halves for each of the three filters. Because the presence of an abrupt turn-on radius is not necessarily realistic for Titan[1], we fixed the turn-on radius and set it very high at $r_1 = 10000$ km (7425 km altitude) to approximate the extinction starting the "top" of the atmosphere. This choice of a very high turn-on radius is further motivated by the lack of an abrupt jump in $\tau_{obs}(r)$ seen from a visual inspection of the data (Fig. 9). Fixed $r_1$ implies $H_{\tau 1}$ should also be fixed, since the latter does not depend on wavelength. We set $H_{\tau 1}$ to be 800 km, based on the results of fits in which $H_{\tau 1}$ was allowed to be a free parameter. The model fit results for $r_2$ (with fixed $H_{\tau 1}$) are shown in Table 4 and may be regarded as quantitatively correct. The error bars on $r_2$ are the formal errors from the least squares fitting and are likely an underestimate given the uncertainty in the choice of $H_{\tau 1}$. For reference, varying $H_{\tau 1}$ from 700 to 1000 km changes $r_2$ by about 15 to 30 km, and this is a more reasonable estimate of the error on $r_2$. Figure 9 shows the model fits for $\tau(r)$ with the appropriate $\tau_{obs}(r)$ data; Figure 10 re-displays all the model curves together to better compare the results at the three filter wavelengths.

<div style="text-align:right">

*\*Insert Table 4*

*\*Insert Fig. 9*

*\*Insert Fig. 10*

</div>

Table 4 and Figure 10 show that the optical depth at a fixed altitude increases with decreasing wavelength, which is consistent with the light curve inversion temperature profiles. Likewise, $r_2$ increases with decreasing wavelength. A notable exception is the *u'* emersion curve, which lies between the *i'* and *g'* curves. This behavior is likely not due to the assumed temperature profile: in an earlier version of this analysis, we used a theoretical temperature profile from Yelle (1991) instead of the HASI profile, yet the $u'$ emersion curve had a similar inconsistency. The $u'$ light curve was more difficult to calibrate because of increased background (c.f. Figs. 2 and 3); thus the discrepancy may be a result of poor calibration.

---

[1] An abrupt turn-on radius is realistic for Pluto, for which this model was originally constructed (EY92).

As stated previously, we derived the extinction parameters in Table 1 by modeling the full light curve and those in Table 4 after assuming a temperature profile and separating out the extinction. The values of $r_2$ in Table 1 are consistently lower than in Table 4 (except for $u'$ emersion), and $r_1$ is much lower in Table 1 than the assumed top of the atmosphere (10000 km) implicit in Table 4. For the method presented in this section (Table 4), $r_2$ did not change within the error bars when $r_1$ was varied, including when $r_1$ was set to the value in Table 1. Although the method presented in the current section requires the assumption of a temperature profile, we are more inclined to choose its results as being more accurate since it decouples the extinction parameters from other light curve parameters.

In order to quantify the dependence of line of sight optical depth on wavelength, we have assumed a separable solution of the form:

$$\tau(r, \Lambda) = h(r) \Lambda^{-p} \tag{17}$$

where $h(r)$ is a function that depends on height only, $\Lambda$ is the wavelength, and $p$ is the power index that describes the wavelength dependence of the extinction. While Eq. 17 is not the same form as Eq. 4, Eq. 17 is the form of assumed by Hubbard et al. (1993), with which we would like to compare values of $p$ (their $q$ parameter). Note that $\tau$ defined in Eq. 4 is implicitly dependent on $\Lambda$ because $r_2$ is a function of wavelength. By considering two different wavelengths, $h(r)$ may be eliminated in Eq. 17, and the resulting equation solved for $p$:

$$p = \log_{10}\left(\frac{\tau(r,\Lambda_1)}{\tau(r,\Lambda_2)}\right) \bigg/ \log_{10}\left(\frac{\Lambda_2}{\Lambda_1}\right). \tag{18}$$

To a very good approximation, $\tau(r,\Lambda_1)/\tau(r,\Lambda_2)$ is independent of $r$ when $\tau$ is specified using Eq. 4 and the model fit values of $r_2$ from above (i.e., Table 4). Thus $p$ is independent of height.

We used Eq. 18 to solve for $p$ six times: by pairing up $i'$ and $g'$, $i'$ and $u'$, and $g'$ and $u'$ once for immersion and once for emersion. Figure 11 shows the results for the individual values of $p$. Because the $u'$ emersion curve did not agree with the others in Fig. 10, $p$ values involving combinations with $u'$ emersion (namely $i'$-$u'$ emersion and $g'$-$u'$ emersion) should be treated with caution. The formal errors on each of the $p$ values, which came from propagating the errors on $r_2$, are also likely underestimates since the errors on $r_2$ were determined to be underestimates as discussed previously. We took the error on the mean value of $p$ to be the standard error of the sample mean. This yields a mean and error for $p$ (excluding $u'$ emersion) of $1.3 \pm 0.2$; c.f. $1.7 \pm 0.2$ obtained by Hubbard et al. (1993).

*<u>*Insert Fig. 11</u>*

As a consistency check, we recalculated $p$ using the extinction parameters found from the light curve fit (i.e., Table 1). Since the light curve model fits for immersion and emersion together, for this calculation we computed $p$ three times: by pairing up $i'$ and $g'$, $i'$ and $u'$, and $g'$ and $u'$. The average of the resulting three values was $1.5 \pm 0.1$, which is consistent with the value stated above.

## 6. Spikes Inversion

The un-averaged light curve shows numerous examples of "spikes," or sharp intensifications in flux above the baseline curve (Fig. 12; Paper I), which can be attributed to focusing by small atmospheric density variations (Brinkmann, 1971; Elliot et al., 1974; Elliot and Veverka, 1976). Because of the dependence of atmospheric refractivity on wavelength, light of different wavelengths arrives at the observer at different times. Spikes are ordered such that the longer wavelength components arrive at larger distances from the center of the shadow plane; hence, spikes arrive in the order of long to short wavelengths during immersion and vice versa during emersion. A set of atmospheric spikes (comprising the three wavelengths) from a single density feature tends to have the same shape, but is displaced in time, which is one way they can be distinguished from noise. Because ULTRACAM uses a different CCD detector for each wavelength, readout noise should not be correlated with wavelength. Photon noise is also not correlated with wavelength. Spikes are not observed in the unocculted part of the light curve, so scintillation effects from Earth's atmosphere can be ruled out. We also assume that fluctuations in the starlight itself are negligible.

****Insert Fig. 12***

In each half of the light curve, we identified over 60 possible pairs of spikes (that is, 60 $g'$ and $i'$ pairs and 60 $u'$ and $i'$ pairs). For each spike, we created a third order polynomial interpolation function to approximate the time dependence of the flux; the function included the spike and a few baseline points on each side. If spikes were close together, the interpolation function included two or three spikes. To find the delay for a given pair (e.g., a matching $g'$ and $i'$ spike), we used a cross-correlation function:

$$F(\sigma'_{gi}) = \int_{t_-}^{t_+} f_i(t) f_g(t + \sigma'_{gi}) dt . \qquad (19)$$

Here $f_g$ is the $g'$ light curve flux (represented by the interpolation function), $f_i$ is the $i'$, $t$ is time, $t_-$ and $t_+$ are the lower and upper time bounds of the interpolation function, and the spike delay $\sigma_{gi}$ is the $\sigma'_{gi}$ for which $F$ is maximum. The delay between $u'$ and $i'$ spikes may be determined by replacing $g$ with $u$ in Eq. 19, as is true for the rest of the equations in this analysis.

Once the spike delay, $\sigma_{gi}$, has been found, it may be related to the $i'$ refraction angle, $\theta_i$, by combining Eqs. 2 and 3 of Elliot et al. (1974), which gives

$$\theta_i = \left(1 - \frac{v_g}{v_i}\right)^{-1} \frac{V}{D} \sigma_{gi}, \tag{20}$$

where $v_g$ and $v_i$ are the atmospheric refractivities evaluated at the wavelengths of the $g'$ and $i'$ filters, respectively; $V$ is the shadow velocity (taken from Table 1); and $D$ is the distance between the observer and the occulting body (taken from Table 2). We adopted the convention that $\sigma_{gi}$ is positive (negative) on immersion (emersion). We assumed the refractivity ratio, $v_g/v_i$, was locally independent of height (i.e., temperature and pressure) in the region of the occultation. As in §4, we assumed an entirely $N_2$ atmosphere, with refractivities determined from the empirical formulas given by Peck and Khanna (1966).

In addition to measuring the spike delays, we also measured the $i'$ spike time $t_i$, which was taken as the mid time between the times of half-maximum flux in the $i'$ spike. The spike time can be related to shadow plane distance, $y_i$, by Eq. 16. To summarize, for each spike pair, we measured $(t_i, \sigma_{gi})$, which are converted to $(y_i, \theta_i)$ using Eqs. 16 and 20. We grouped the measurements by immersion and emersion. Similarly, we measured $(t_i, \sigma_{ui})$ for the $u'$ and $i'$ spike pairs (where $\sigma_{ui}$ is the delay between the $u'$ and $i'$ spikes), converted to $(y_i, \theta_i)$, and grouped the measurements by immersion and emersion. Table 5 shows the number of spikes in each group. The number of spike sets is less than the value of 60 mentioned above because a maximum in the cross-correlation function (Eq. 19) and hence $\sigma_{gi}$ or $\sigma_{ui}$ could not always be found.

*Insert Table 5*

The sets of $(y_i, \theta_i)$ determined from the spikes analysis may be inverted to derive temperature as a function of radius from the body's center (easily converted to altitude) using several equations from the light curve inversion procedure of Elliot et al. (2003) that was described in §4. The spikes inversion method uses these equations in a novel way. We state the relevant equations in detail here to illustrate this new method, but leave their derivations to Elliot et al. (2003). The light curve inversion procedure divides the light curve into two regions: the boundary region and the inversion region. Points in the boundary region are least squares fit to a light curve model to determine an upper boundary condition. The temperature at the $l+1/2$ th radius shell[2] in the inversion region is given by

$$T(r_b, r_{l+1/2}) = \frac{\mu m_{amu} GM_p \left[ B_p(r_b, r_{l+1/2}) + S_p(r_b, r_{l+1/2}) \right]}{k \left[ B_\nu(r_b, r_{l+1/2}) + S_\nu(r_b, r_{l+1/2}) \right]} \tag{21}$$

where $r_b$ is the boundary radius, $B_\nu$ is the refractivity boundary integral, $B_p$ is the pressure boundary integral, $S_\nu$ is the refractivity summation term and $S_p$ is the pressure summation term. We used the same upper boundary conditions (i.e. the same values of $r_b$, $B_\nu$, and $B_p$) as in the $i'$ light curve inversion (§4) because the $i'$ light curve contained the least extinction, which was not accounted for in the determination of the boundary conditions. The $i'$ immersion half of the light curve was used in deriving the boundary condition for the spikes that occurred during immersion, and the *i'* emersion half of the light curve was used for the emersion spikes.

---
[2] We have switched the index $i$ used by Elliot et al. (2003) to $l$ to avoid confusion with the $i$ subscript used in this paper to denote $i'$ light curve variables.

The summation terms are specified by[3]

$$S_v(r_b, r_{l+1/2}) = -\frac{1}{\pi} \sum_{j=l_b}^{l} \frac{(\theta_{j+1/2} - \theta_{j-1/2})}{z_{j+} - z_{j-}} \left[ z_{j+} \cosh^{-1} z_{j+} - \sqrt{z_{j+}^2 - 1} - z_{j-} \cosh^{-1} z_{j-} + \sqrt{z_{j-}^2 - 1} \right]$$

(22)

and

$$S_p(r_b, r_{l+1/2}) = -\frac{1}{\pi\, r_{l+1/2}} \sum_{j=l_b}^{l} \frac{(\theta_{j+1/2} - \theta_{j-1/2})}{z_{j+} - z_{j-}} \left[ z_{j+} \cosh^{-1} z_{j+} - 2\sqrt{z_{j+}^2 - 1} - \sin^{-1}\left(\frac{1}{z_{j+}}\right) \right.$$

$$\left. - z_{j-} \cosh^{-1} z_{j-} + 2\sqrt{z_{j-}^2 - 1} + \sin^{-1}\left(\frac{1}{z_{j-}}\right) \right],$$

(23)

where $z_{j+}$ and $z_{j-}$ are defined by

$$z_{j+} \equiv \frac{r_{j+1/2}}{r_{l+1/2}}$$

(24)

and

$$z_{j-} \equiv \frac{r_{j-1/2}}{r_{l+1/2}}.$$

(25)

The index $l_b$ is the index that divides the boundary and inversion regions such that $r_{l_b-1/2} = r_b$, and $r_{l_b+1/2}, r_{l_b+3/2}, r_{l_b+5/2}, \ldots$ are the radii in the inversion region in order of higher radii to lower radii. Likewise, $\theta_{l_b+1/2}, \theta_{l_b+3/2}, \theta_{l_b+5/2}, \ldots$ are the corresponding refraction angles, and the boundary refraction angle $\theta_b = \theta_{l_b-1/2_b}$ is specified by the boundary condition. To make use of the spikes

---

[3] Two errors in Eq. 53 of Elliot et al. (2003) have been corrected in Eq. 23, namely, the omission of a $1/r_{l+1/2}$ and the replacement of $\sin^{-1} z$ with $\sin^{-1}(1/z)$.

data in Eqs. 21–25, the discrete version of Eq. 15 is used:

$$r_{l+1/2} = y_{l+1/2} - D\theta_{l+1/2}. \tag{26}$$

Eqs. 21–26 provide a method of determining $T(r)$ from a set of $(y,\theta)$ data. The difference between the light curve inversion and the spikes inversion is that in the former case, $(y,\theta)$ is determined from light curve flux as a function of time, but in the latter $(y,\theta)$ is determined from the spike delay between two wavelengths as a function of time.

Figure 13 shows the results from inverting the spikes data. Only spikes whose corresponding radii lie in the boundary region could be considered in the inversion; hence there are fewer points than the number of spikes listed in Table 5. We have not yet developed a formal procedure for determining the errors on the spikes inversion temperature. We did however investigate the effect of changing the boundary flux level from 0.5 to 0.4 and 0.6. Higher (lower) fluxes correspond to higher (lower) altitudes; thus raising (lowering) the boundary level flux raises (lowers) the boundary radius, and the temperature profile extends to higher (lower) altitudes. As is generally the case in occultation temperature inversions, the range of temperature values at a given altitude is large at higher altitudes and converges at lower altitudes, where the choice of boundary condition becomes less important.

*Insert Fig. 13*

The "spikes inversion" temperatures represent the true atmospheric temperature, since extinction does not directly affect the spike delays. Extinction could be affecting the boundary condition in both the spikes inversion and the light curve inversion. It would be possible to derive a boundary condition from the light curve with a light curve model that contained extinction, from the HASI temperature profile, or from a theoretical temperature model (e.g., Yelle, 1991). An investigation of the inversion profiles resulting from these other boundary conditions is beyond the scope of this paper.

The spikes inversion temperature profiles derived with a boundary flux level of 0.5 (black points in Fig. 13) and the $i'$ light curve inversion temperature profiles (also shown in Fig. 13), which share the same boundary condition, are consistent: the latter are a lower limit on temperature because of extinction in the light curve, and are indeed less than the former. From the $g'$ and $i'$ (and perhaps the $u'$ and $i'$ immersion) spikes inversion temperature profiles (Fig. 13), the temperature increases with decreasing altitude. This trend is consistent with the HASI temperature profile (dotted line in Fig. 13), but the HASI profile is more isothermal. The spikes inversion temperatures are equal to or warmer than the HASI data. Since we have yet to derive a formal procedure for estimating the errors on the spike inversion temperatures, it is not possible to quantitatively assess how well they agree with the HASI data. Qualitatively, the spikes inversion temperatures are within reasonable range of the HASI data, and further use and development of the spikes inversion method is warranted.

## 7. Central Flash

We observed an unusual, three-peaked central flash in $i'$ filter (Fig. 14) due to the close proximity of the star's path to the center of Titan's disk. In general, the central flash occurs near the mid time of an occultation, when the second term in parenthesis of Eq. 10 (the focusing term) becomes large (see Elliot and Olkin, 1996). The focusing, and hence the shape of the central flash, depends heavily on the occulting body's shape. For example, a body with a circular shape produces one peak if the chord passes over the exact center; a body with an elliptical shape produces two or four peaks if the chord is near the center (Elliot et al., 1977). No peak is observed when the chord is too far from the center, or atmospheric extinction renders it undetectable (such as in our $g'$ and $u'$ light curves).

As one might expect, our three-peaked central flash does not agree well with a circular or elliptical light curve model. This disagreement was also noted by Hubbard et al. (1993) in their 28 Sgr occultation light curves. They utilized a shape model that was the sum of Legendre polynomials:

$$r_3 = b[1 + \sum_{l=1}^{\infty} f_{2l}(P_{2l}(\cos\theta_3) - 1)].  \tag{27}$$

Here $b = 3062.5$ km is the polar radius, $f_{2l}$ is a shape coefficient, $P_{2l}$ is the $2l$ th Legendre polynomial, and $\theta_3$ is the colatitude measured from the (north) polar axis. Their nonuniform, deprojected solution (Table 4 of Hubbard et al., 1993) yielded $f_2 = -0.00445$, $f_4 = -0.00363$, $f_6 = -0.00216$, and $f_8 = -0.00067$ (they truncated higher order terms).

After projecting this shape to the particular geometry of our occultation and calculating the resulting $i'$ light curve from the atmospheric parameters of our circular model (i.e., Table 1)[4], we find a central flash with three major peaks (Fig. 14). Our observed $i'$ light curve does not closely match any of those observed by Hubbard et al. (1993), yet their shape model nearly agrees with our data (in terms of the number of peaks and their timings). This is a confirmation of the validity of their shape model, especially notable because none of their light curves had three major peaks. It was necessary to use a different value of the distance of closest approach $y_{min}$ (19 km) to obtain a matching model light curve, but this is a minor concern since the value of $y_{min}$ derived from the circular model will be only an approximation (the fit value of $y_{min}$ is sensitive to the central flash region, where the circular model is inaccurate). Shifting the time of closest approach by +0.2 s also gave better agreement; such a shift is within the error bars of the time of closest approach found in the circular model (Table 1).

---

[4] Atmospheric parameters from the circular model may be used since these do not depend greatly on the shape of the atmosphere.

The model light curve underestimates the peaks' amplitudes. This may be due to the model extinction being too large at this altitude. The extinction model increases exponentially as altitude decreases; thus extrapolation to the lower altitudes of the central flash region may be overestimating the optical depth. Reducing the optical depth by a factor of 0.9 yields better agreement. The light curve model also produced two small peaks within the two outer major peaks that do not appear in the data; but noise in the data may have prevented them from being clearly resolved. It is also possible that the two small peaks are numerical errors.

## 8. Discussion

The inversion temperatures derived in §4 and §6 are of importance since few observations exist for this region of Titan's atmosphere. The widely-observed occultation of 28 Sgr produced several light curve inversion temperature profiles (Hubbard et al., 1993). The resulting inversion temperatures were generally between 150 and 170 K (between 300 and 500 km altitude). No significant difference existed between immersion and emersion. Hubbard et al. (1993) observed a pronounced downturn in inversion temperature with decreasing altitude below 380 km, which they attributed to increasing haze optical depth (and therefore the inversion temperatures were not equivalent to the true temperature). Our light curve inversion temperatures also show this downturn, which occurs around 400 km in the $i'$ profiles, 425 km in the $g'$ profiles, and 450 km in the $u'$ immersion profile (we neglect the $u'$ emersion profile in this discussion having shown this part of the light curve to be inconsistent with the other parts).

Above the downturn, the light curve inversion temperature is nearly isothermal, with values of 165 ± 10 K at $i'$, 150 ± 10 K at $g'$, and 125 ± 15 K at $u'$ immersion. While the $i'$ and $g'$ light curve inversion temperatures agree with the 28 Sgr occultation results, the $u'$ immersion is colder.

All of these light curve inversion temperatures are the same as or colder than other measurements including: Yelle (1991), whose model predicts 160–170 K between 325–500 km; Griffith et al. (2005), who obtained 180 ± 2 K between 230 ± 20 to $380^{+50}_{-100}$ km using $CH_4$ spectra (assuming an isothermal atmosphere between the given altitudes), Tracadas et al. (2001), who obtained 180 ± 30 K from a stellar occultation (assuming an isothermal atmosphere); Fulchignoni et al. (2005), who from HASI data obtained monotonically decreasing temperatures from 185 K at 250 km to 150 K at 500 km (Fig. 7); Vinatier et al. (2007), who from Cassini CIRS data at 13°S latitude obtain decreasing temperature from 183 K at 310 km to 160 K at 500 km; and Shemansky et al. (2005), who from Cassini UVIS data derive a temperature of about 175 K at 450 km, which is the lowest altitude this data covers. The colder inversion temperatures are consistent with the postulate that light curve inversion temperature is a lower limit on actual temperature due to the presence of extinction.

One possible reason for the differences between our data and the various measurements themselves could be that all of these data were taken at different seasons and latitudes or are disk-integrated in the case of Griffith et al. (2005). Given that extinction by haze causes both colder and wavelength-varying light curve inversion temperatures, our light curve inversion temperatures may also be affected by extinction above the downturn altitude mentioned above. The Imaging Science Subsystem (ISS) on the Cassini spacecraft observed haze as high as 500 km (Porco et al., 2005). Hubbard et al. (1993) suggest that there are two haze components: a high-altitude component and a low-altitude component. It is plausible that the low-altitude haze component produces the downturn in inversion temperature at lower altitudes, and the high-altitude haze component causes the light curve inversion temperature to be colder than the true temperature at higher altitudes.

The spikes inversion temperatures (Fig. 13) are approximately equal to or warmer than the HASI data and span the range of 147–207 K. This range includes the previous measurements listed above. The $g'$ and $i'$ immersion profiles of varying boundary level flux appear to be converging to a temperature of 195 K at 395 km. This temperature is higher than previous measurements except for Tracadas et al. (2001). Again we note that the spikes inversion temperatures are higher than the light curve inversion temperatures. The latter is a lower limit on temperature while the former is the actual temperature, thus they are not inconsistent.

We were unable to resolve the distinct haze layers observed by the Cassini ISS (Porco et al., 2005). The highest haze layer it detected was near 500 km altitude. According to our extinction analysis (Fig. 10), the optical depth at this altitude is 0.1–0.3. We determined the power index of the wavelength dependent component of the haze extinction (§5) to be $1.3 \pm 0.2$, which is slightly lower than the value obtained by Hubbard et al. (1993) of $1.7 \pm 0.2$. Their value is however within the scatter of our data points in Fig. 11. Tomasko et al. (2005) report properties of the haze extinction determined from the Huygens Probe Descent Imager/Spectral Radiometer instrument. Their Fig. 21 implies a power index of 2. However, this value was derived from observations taken much closer to the surface.

The extinction model used in our analysis was adapted from a Pluto model, and the presence of an abrupt haze turn-on radius as a model parameter is unrealistic for Titan. A future analysis could utilize a more appropriate extinction model. Hubbard et al. (1993) modeled Titan's atmospheric haze with two layers: a high altitude layer that was confined to certain latitudes and a low altitude layer that existed everywhere. An even more realistic haze model would include parameters that describe the distinct haze layers observed by the Cassini ISS (Porco et al., 2005), as well as any latitudinal or seasonal dependencies that might exist.

## 9. Conclusions

We have presented the results from an analysis of the light curves from the 2003 Nov. 14 occultation by Titan that occurred just before 7:00 UT. The half-light radius of this occultation was lower than those reported for previous occultations. Inversion of the light curves yielded a lower limit on temperature in the region between 335 and 485 km altitude. The light curve inversion profiles agree with previous models and observations at higher altitudes, but diverge towards colder temperatures at lower altitudes. This divergence is a sign of wavelength dependent extinction that increases with decreasing altitude and wavelength. After assuming a temperature profile, we separated out the extinction effects from the light curve and quantified their vertical structure using a model from EY92. We presented a new method for determining temperature from scintillation spikes in the light curve. This method was subject to the specific boundary condition used, but yielded inversion temperatures that were in agreement with or warmer than previous measurements.

We found the three-peaked structure of the central flash in the $i'$ curve to be particularly intriguing. Paper I also suggests that two minor peaks lie between the major peaks. To our knowledge, this structure has not been observed in any planetary occultation. The shape coefficients derived by Hubbard et al. (1993) from the 28 Sgr occultation produce a three-peaked central flash that match our observed central flash reasonably well, despite none of their light curves exhibiting the same structure as ours. It would be interesting in a future analysis to derive shape coefficients from our data as a comparison and consequently derive the zonal stratospheric winds.

## Acknowledgements

A. Zalucha gratefully acknowledges support from NSF grant AST-0073447 and NASA grant NNG04GF25G. ULTRACAM is supported by PPARC grant PP/D002370/1. We thank Lawrence H. Wasserman and an anonymous referee for their helpful comments.

Note: During the refereeing process for this paper, Sicardy et al. (2006) published results for two occultations by Titan, including one that is the subject of this paper.

**Tables**

TABLE 1

RESULTS FROM LIGHT CURVE MODEL FIT

|  | Filter | | |
|---|---|---|---|
| Fixed parameters | $i'$ | $g'$ | $u'$ |
| Number of points fit (after averaging) | 598 | 598 | 598 |
| Shadow velocity[a], $V$ (km s$^{-1}$) | 11.25 | 11.25 | 11.25 |
| Integration time, $\Delta t$ (s) | 2 | 2 | 2 |
| Full-scale level, $s_f$ | 1 | 1 | 1 |
| Thermal gradient parameter, $b$ | –1.5 | –1.5 | –1.5 |
| | | | |
| Fitted parameters | | | |
| Background fraction, $s_{bf}$ ($10^{-2}$) | –1.2 ± 0.3 | 0.5 ± 0.2 | 16.6 ± 0.4 |
| Background slope, $s'$ ($10^{-5}$ s$^{-1}$) | 1.8 ± 0.3 | 1.5 ± 0.2 | 13.4 ± 0.5 |
| Half-light radius[b], $r_H$ (km) | 2992 ± 13 | 2967 ± 10 | 2971 ± 23 |
| Half-light equivalent isothermal lambda, $\lambda_{Hi}$ | 51 ± 3 | 56 ± 2 | 62 ± 11 |
| Radius of extinction onset, $r_1$ (km) | 3173 ± 22 | 3194 ± 9 | 3214 ± 20 |
| Radius of optical depth unity, $r_2$ (km) | 2868 ± 5 | 2923 ± 4 | 2956 ± 11 |
| Haze scale height, $H_{\tau 1}$ (km) | 90 ± 15 | 97 ± 8 | 97 ± 20 |
| Time of closest approach, $t_{mid}$ (seconds after 2003 11 14 06:00:00 UT) | 3236.1 ± 0.2 | 3236.3 ± 0.1 | 3236.7 ± 0.3 |
| Distance of closest approach[c], $y_{min}$ (km) | 39 ± 8 | 39 | 39 |
| | | | |
| Derived Parameters | | | |
| Temperature at half-light[d] (K) | 185 ± 11 | 172 ± 6 | 156 ± 27 |
| Pressure scale height at half-light[e] (km) | 55 ± 3 | 50 ± 2 | 45 ± 8 |
| Half-light altitude (km) | 417 ± 13 | 392 ± 10 | 396 ± 23 |
| Altitude of extinction onset (km) | 598 ± 22 | 619 ± 9 | 639 ± 20 |
| Altitude of optical depth unity (km) | 293 ± 5 | 348 ± 4 | 381 ± 11 |

[a]Velocity near the occultation mid time. The respective velocities at the beginning and end of the occultation were 11.26 km s$^{-1}$ and 11.23 km s$^{-1}$.
[b]The surface radius of Titan is 2575 km.
[c]Fit for $i'$ only; n.b. the central-flash analysis yields a closest approach distance of 19 km.
[d]Given by $\mu m_{amu} G M_p / k(\lambda_{Hi} - 5b/2) r_H$ ; see Table 2 for values of $\mu$ and $M_p$.
[e]Given by $r_H / (\lambda_{Hi} - 5b/2)$.

TABLE 2

LIGHT CURVE INVERSION PARAMETERS

| Parameter | Value |
|---|---|
| Mass[a], $M_p$ (kg) | $1.3454 \times 10^{23}$ |
| Distance[b], $D$ (AU) | 8.38 |
| Atmospheric composition | $N_2$ |
| $i'$ refractivity at STP[c] | 0.00234841 |
| $g'$ refractivity at STP[c] | 0.00238879 |
| $u'$ refractivity at STP[c] | 0.00244988 |
| Mean molecular weight, $\mu$ (amu) | 28.01 |
| Boundary flux level | 0.5 |
| Minimum shell thickness | 1.0 km |

[a]Jacobson et al. (2005).
[b]Jacobson (1996).
[c]Peck and Khanna (1966).

TABLE 3

DISTRIBUTION OF POINTS IN THE INVERSION CALCULATION

| Inversion | Number of boundary points | Number of inversion points |
|---|---|---|
| $i'$ immersion | 1111 | 600 |
| $g'$ immersion | 1111 | 304 |
| $u'$ immersion | 1111 | 103 |
| $i'$ emersion | 895 | 670 |
| $g'$ emersion | 895 | 274 |
| $u'$ emersion | 895 | 173 |

TABLE 4

EXTINCTION MODEL RESULTS

| Light curve part | Radius of optical depth unity, $r_2$ (km)[a] | Altitude of optical depth unity (km) |
|---|---|---|
| $i'$ immersion | $2912 \pm 2$ | $337 \pm 2$ |
| $g'$ immersion | $2936 \pm 2$ | $361 \pm 2$ |
| $u'$ immersion | $2976 \pm 3$ | $401 \pm 3$ |
| $i'$ emersion | $2902 \pm 2$ | $327 \pm 2$ |
| $g'$ emersion | $2952 \pm 2$ | $377 \pm 2$ |
| $u'$ emersion | $2927 \pm 3$ | $352 \pm 3$ |

[a]Optical depth refers to the optical depth along the line of sight, tangent to the limb of Titan.

TABLE 5

DISTRIBUTION OF SPIKES DATA

| Spike set | Number of spike pairs |
|---|---|
| $g'$ - $i'$ immersion | 40 |
| $g'$ - $i'$ emersion | 27 |
| $u'$ - $i'$ immersion | 19 |
| $u'$ - $i'$ emersion | 6 |

**Figure Captions**

Figure 1. Path of occultation star around Titan's limb as seen from Earth (heavy solid line and points). The star's path is in time increments of 1 s, showing that the star's image moves quickly around the southern part of the limb. The directions shown are directions in Earth's sky; the visible pole is the south pole. The edge of the globe is Titan's surface, while the dotted circle is the half-light radius from the $i'$ light curve fit (Table 1). Immersion occurs on the right (west) side, while emersion occurs on the left (east) side. The star's path intersects Titan at nearly the same latitude during immersion and emersion (2°S and 1°N, respectively). The corresponding longitudes are 155°W and 24°E, respectively.

Figure 2. Background-uncorrected light curves (points) for wavelengths $i'$ (top), $g'$ (middle), and $u'$ (bottom). The dashed line indicates the zero flux level; the solid line (mostly obscured by the data points) is the model fit. The data are averaged by 60 points (2 s), effectively averaging out the spikes. The central flash is clearly visible in the $i'$ curve. The $u'$ shows a large linear drift in the background signal.

Figure 3. Background-corrected light curves (points) for wavelengths $i'$ (top), $g'$ (middle), and $u'$ (bottom). The dashed line indicates the zero flux level; the solid line (mostly obscured by the data points) is the model fit now with the background subtracted out. The data are averaged by 60 points (2 s), effectively averaging out the spikes. The correction to the $i'$ and $g'$ is very small.

Figure 4.  Lower limiting temperature profiles obtained from the light curve inversion for wavelengths *i*' (top), *g*' (middle), and *u*' (bottom).  The left column of panels is immersion; the right column of panels is emersion.  The arrows indicate that profiles are a lower limit on temperature.  The error bars are the formal error from the least squares fitting procedure used to obtain the boundary condition.

Figure 5.  Lower limiting temperature profiles obtained from the light curve inversion.  The abbreviations "im" and "em" refer to immersion and emersion respectively.  We also include the temperature profile from HASI over this altitude range (Fulchignoni et al., 2005).  We omit the error bars on the inversion profiles for clarity; the arrows indicate that the light curve inversion profiles are a lower limit on temperature.  The curves all have roughly the same shape but are displaced in temperature, which is evidence for wavelength-dependent extinction (see text).

Figure 6.  Diagram of the adopted coordinate system (adapted from Elliot et al. 2003).  Starlight enters the body's atmosphere, is refracted in the body plane, and reaches the observer in the shadow plane.  The $r$ coordinate is a radial coordinate from the body's center, the $y$ coordinate is measured from the center of the shadow plane, $\theta$ is the refraction angle, $x$ is measured along the ray path, $r'$ is a radial coordinate and variable of integration, and $D$ is the distance to the observer.

Figure 7.  HASI temperature profile from Fulchignoni et al. (2005; points).  The solid line is a least squares fit to the data found by fitting the amplitudes of a Fourier series with 40 frequencies.  The thick vertical bar shows the altitude range of the light curve inversion profiles

as shown in Fig. 5. We did not include measurements below 157 km since the occultation does not probe that deeply into the atmosphere.

Figure 8. Extinction-free light curves generated with HASI data (solid lines). The top panel is $i'$, the middle is $g'$, and the bottom is $u'$. Also shown are the model fits to the observed light curve from Fig. 3 (dashed lines). The fluctuations in the extinction-free light curve are a direct result of the fluctuations in the temperature profile. The extinction-free light curves have higher flux since extinction diminishes the flux in the observed light curves.

Figure 9. Line of sight optical depth data (points) found by dividing the measured light curves with the light curves generated from HASI temperature data. Also shown are the model fits (solid lines) to Eq. 4. The model was fit to the entire data range (252–1380 km altitude), but only data below 800 km is shown. The top panels are $i'$, the middle are $g'$, and the bottom are $u'$; the left panels are immersion and the right are emersion.

Figure 10. Modeled line of sight optical depth as a function of altitude. The abbreviations "im" and "em" refer to immersion and emersion respectively. At fixed altitude the optical depth increases with decreasing wavelength. The $u'$ emersion curve does not behave ideally in that it lies between the $i'$ and $g'$ curves. The disagreement possibly stems from an incorrect calibration of the $u'$ light curve.

Figure 11. Results for the $p$ power index. On the horizontal axis, "i", "g", and "u" denote the different filters while "im" and "em" respectively refer to immersion and emersion. The pairs

involving *u'* emersion (open squares) should be treated with caution since that curve exhibited a different behavior than the others in Fig. 10. The error bars are the formal errors from propagating the errors on $r_2$ and are underestimates (see text). The formal error bars on the *g'* and *i'* pairs are too small to be seen in this plot.

Figure 12. An example of four scintillation spikes during emersion (adapted from Paper I). The arrival of different wavelengths within each set of spikes is ordered *u'* (dotted), *g'* (dashed), *i'* (solid). The quantities of interest are the time delay between the arrival of the *g'* and *i'* spikes, time delay between the arrival of the *u'* and *i'* spikes, and the arrival time of the *i'* spike.

Figure 13. Spike inversion temperature results with different boundary flux levels (squares). Also shown are the HASI temperature profile (dotted line) and the immersion (solid line) and emersion (dashed line) *i'* light curve inversion results of §4. As in Figs. 4 and 5, the arrows indicate that the light curve inversion profiles are a lower limit on temperature. The upper left panel is the result of pairing immersion *g'* and *i'* spikes; the upper right is the same but for emersion. The lower left panel is the result of pairing immersion *u'* and *i'* spikes; the lower right is the same but for emersion. The spike inversion temperatures are approximately equal to or warmer than the HASI temperature profile. They are greater than the light curve inversion temperatures, which is consistent with the latter being a lower limit on temperature.

Figure 14. Central flash region of the *i'* light curve shown at full resolution (points). The dashed line is the light curve constructed using the shape model of Hubbard et al. (1993) and the atmospheric parameters found from the circular light curve model (Table 1) except with $y_{min} =$

19 km and the time of closest approach shifted by +0.2 s. The solid line is the same light curve but with the optical depth reduced by a factor of 0.9. The fact that there are three major peaks at nearly the correct times and amplitudes is a remarkable confirmation of the validity of the shape model of Hubbard et al. (1993).

**Figures**

A. Zalucha, Figure 1

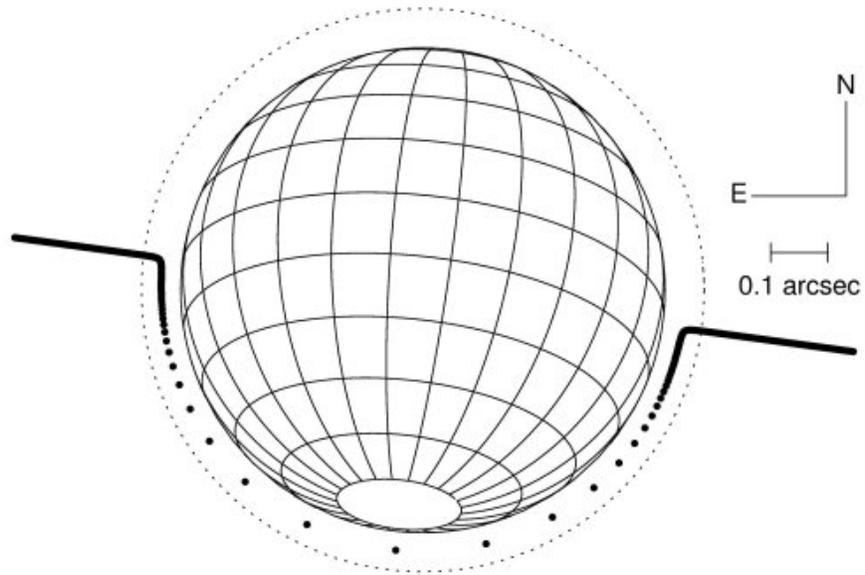

A. Zalucha, Figure 2

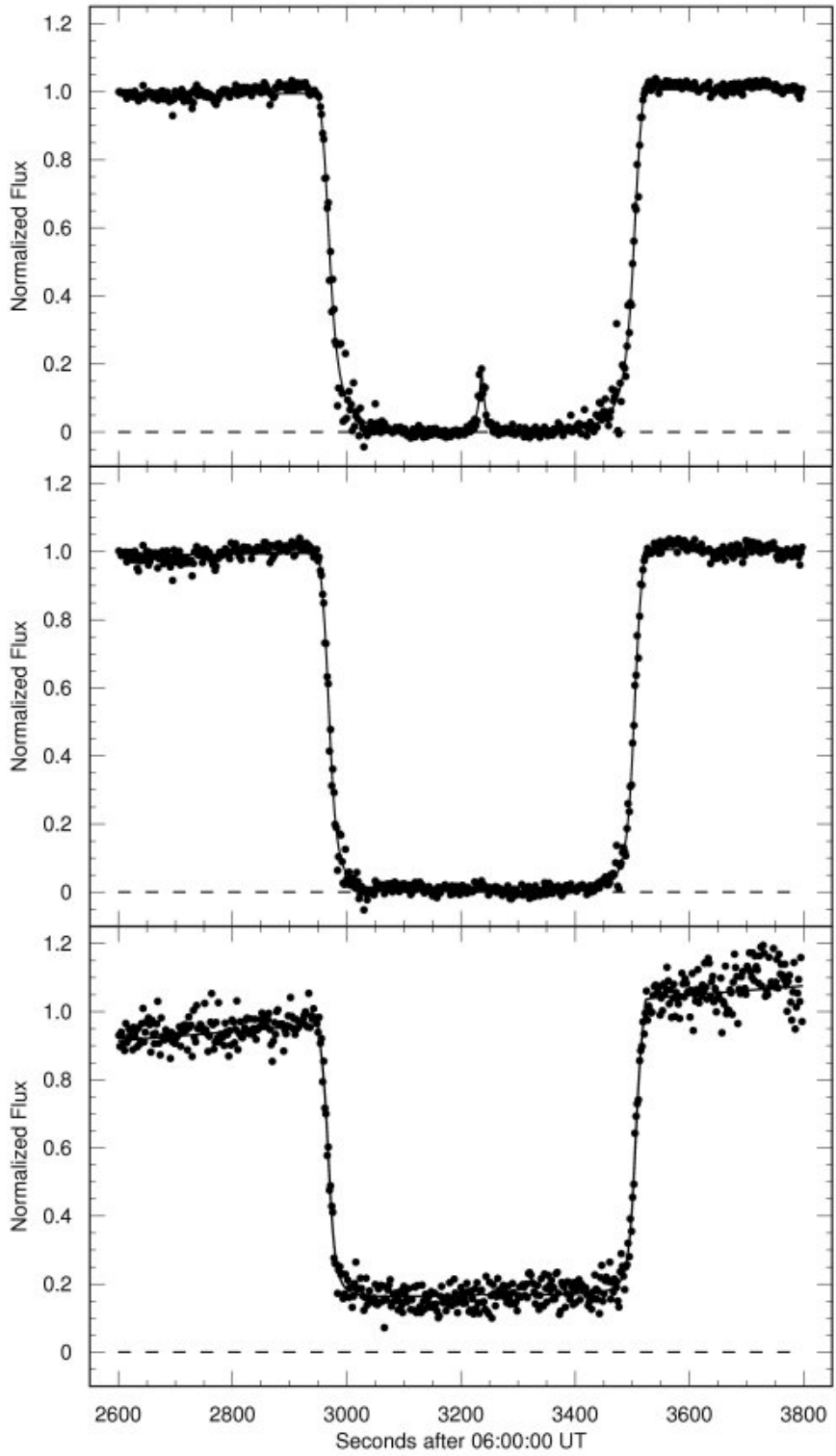

A. Zalucha, Figure 3

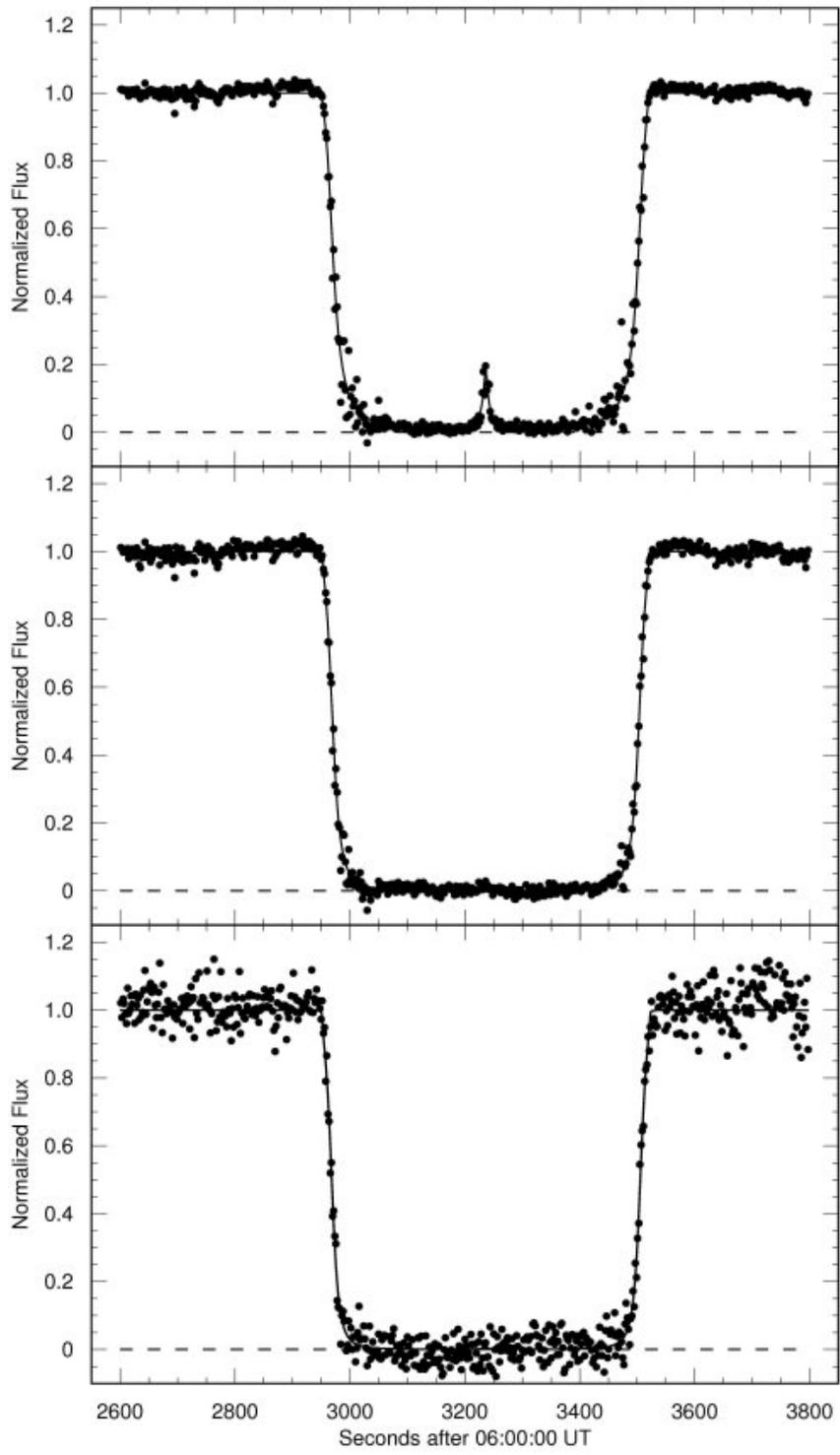

A. Zalucha, Figure 4

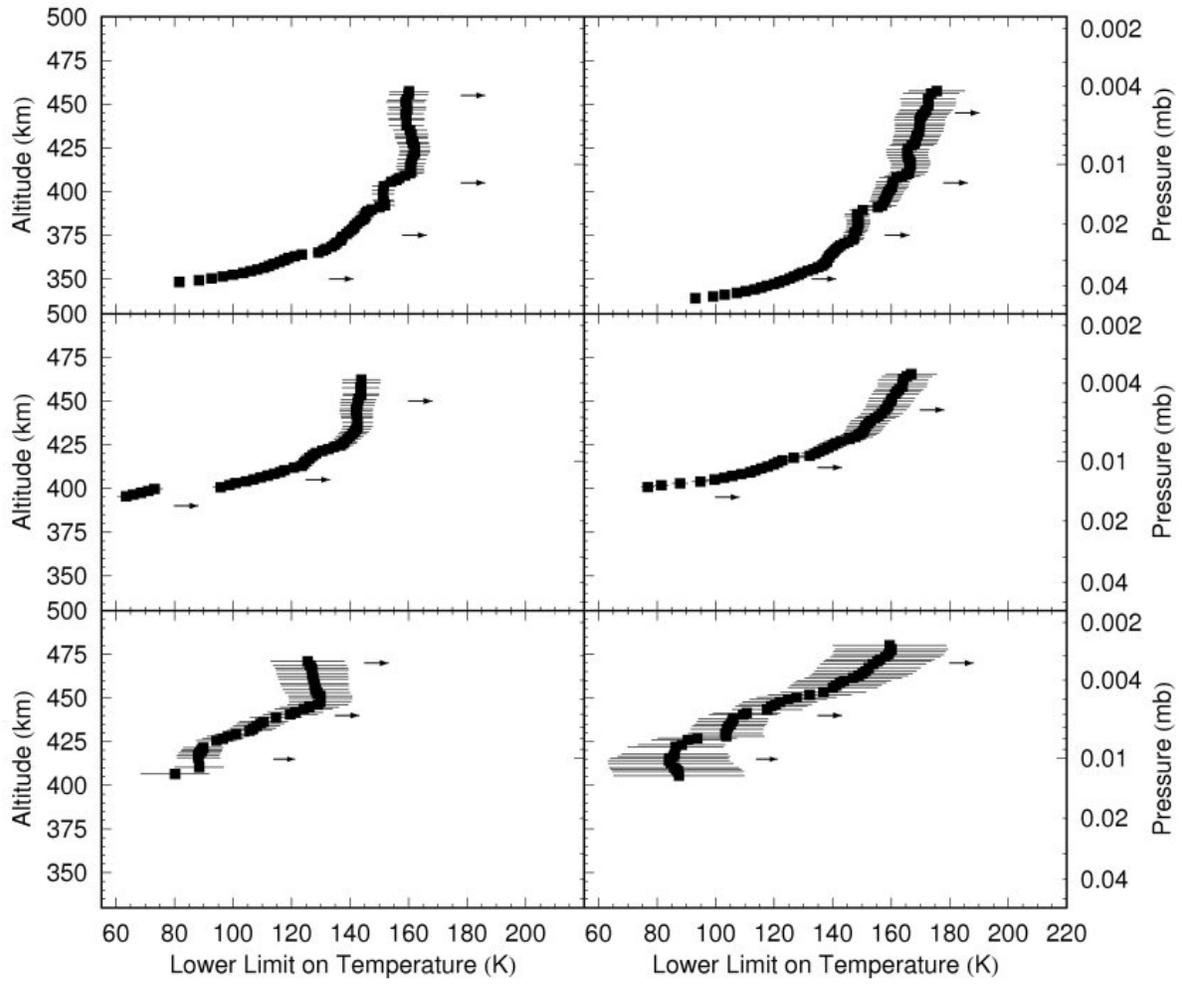

A. Zalucha, Figure 5

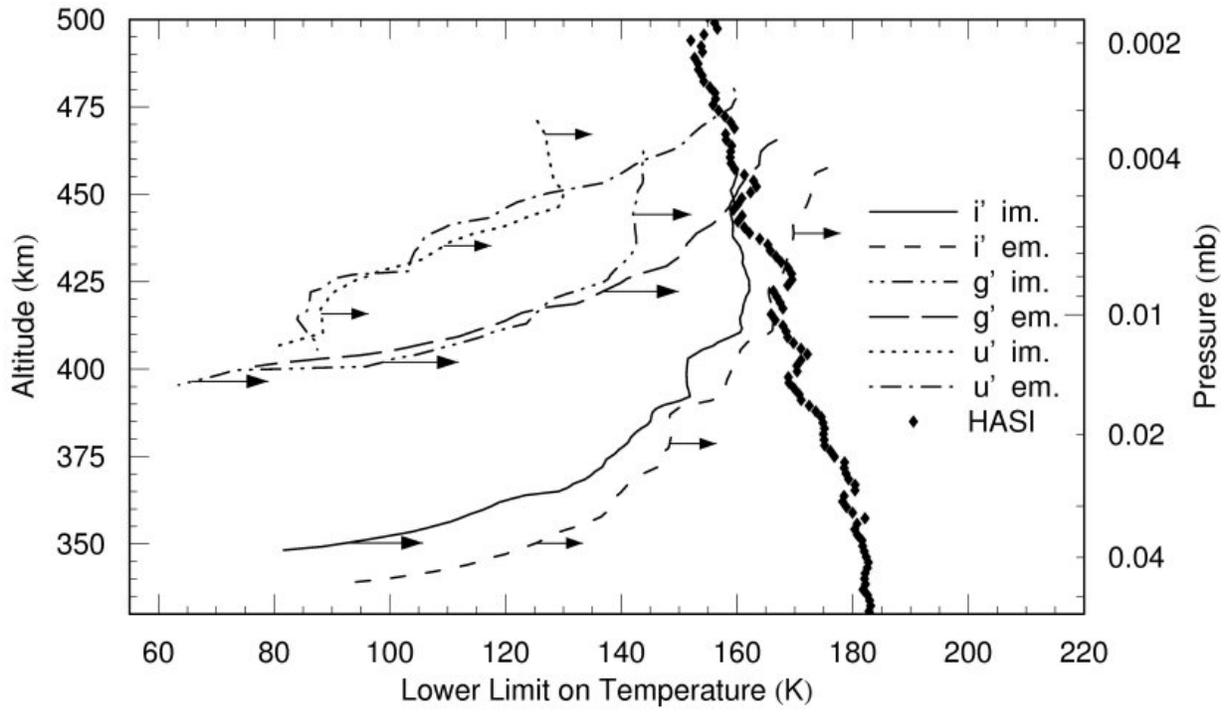

A. Zalucha, Figure 6

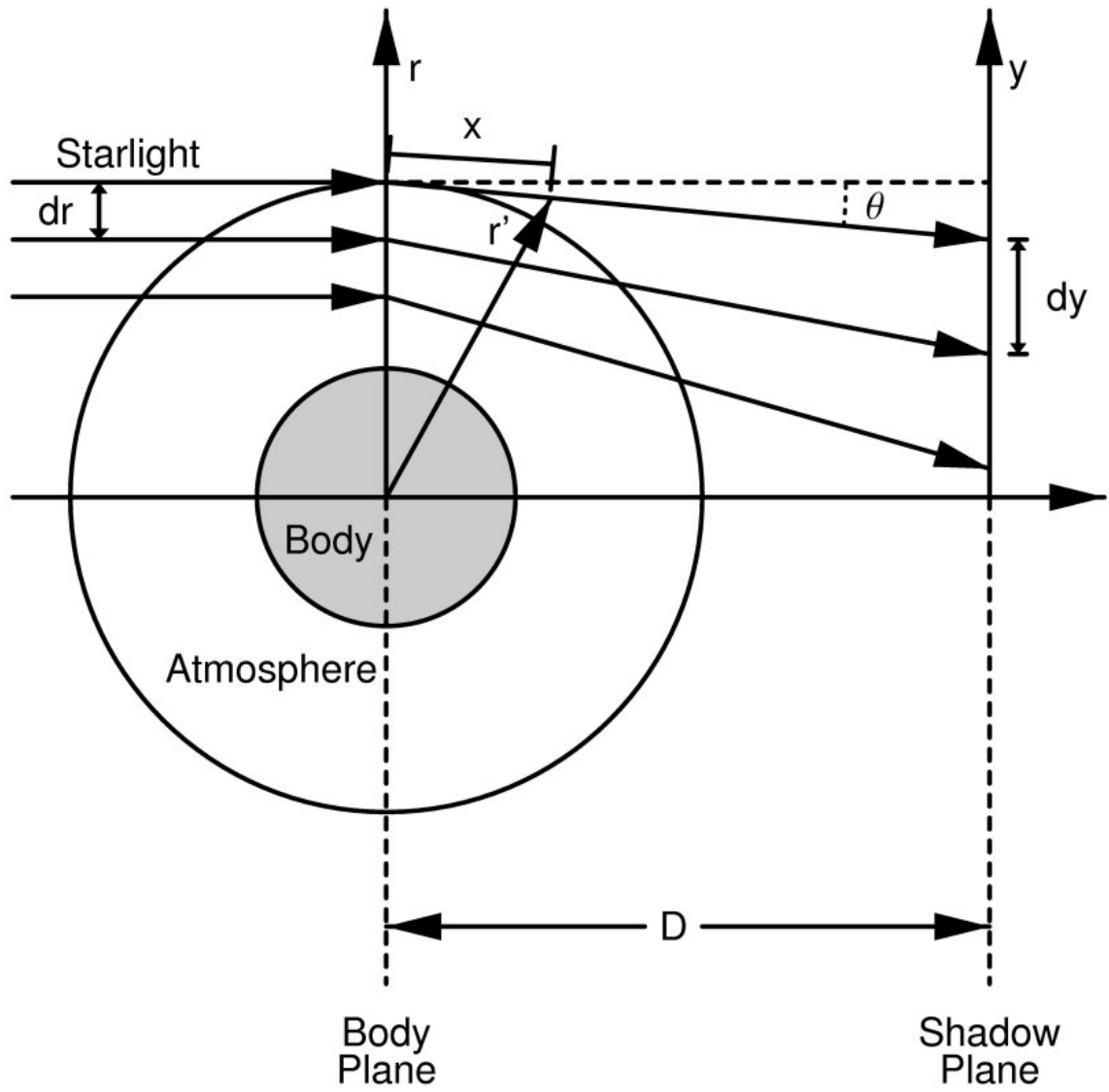

A. Zalucha, Figure 7

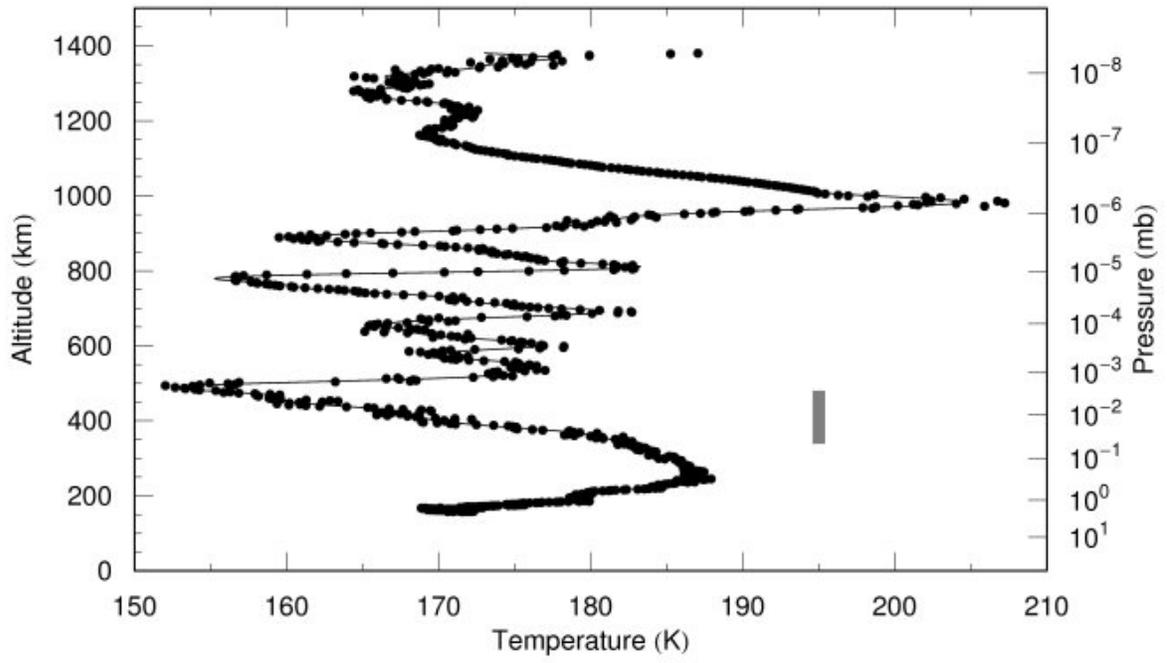

A. Zalucha, Figure 8

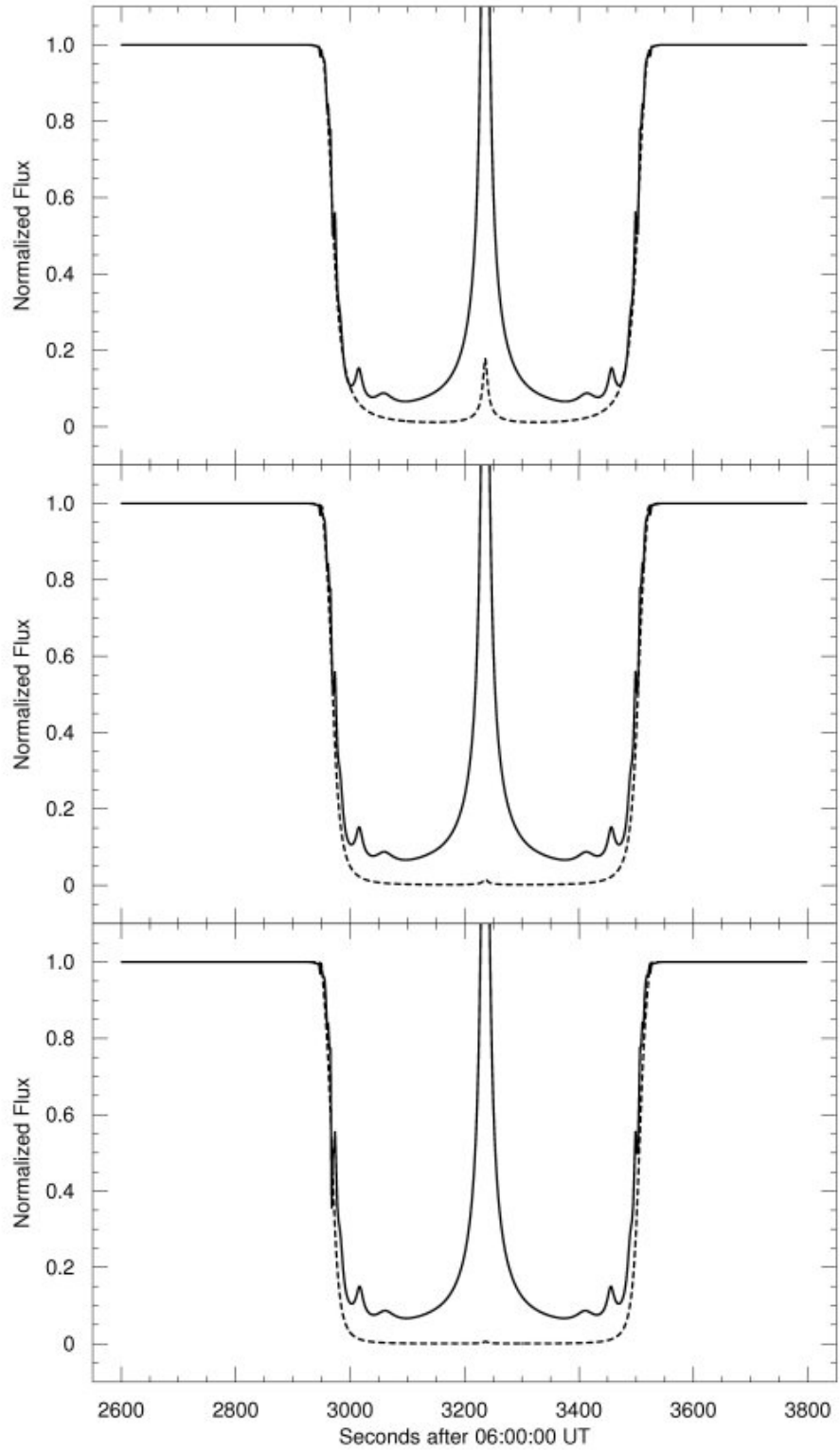

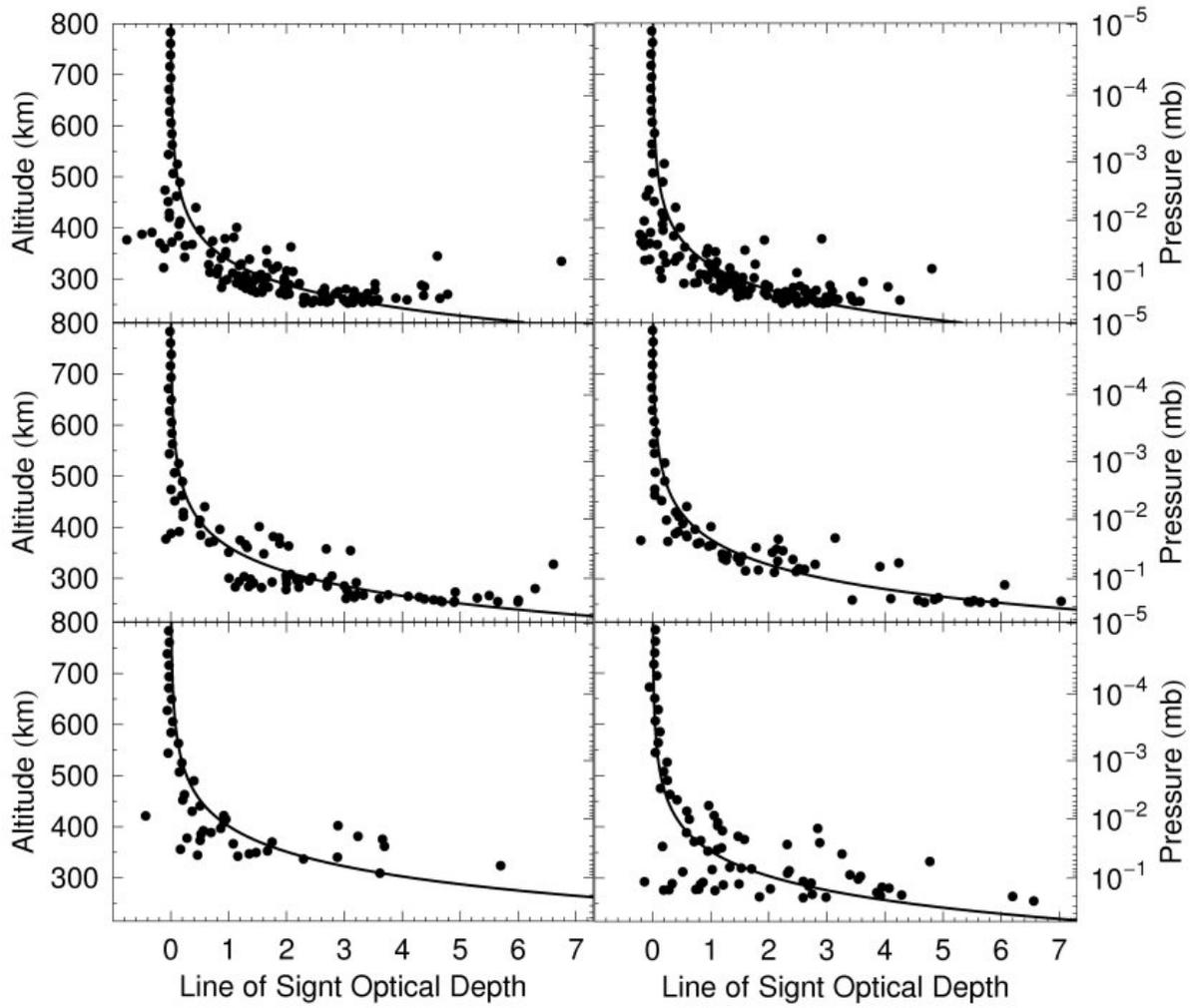

A. Zalucha, Figure 10

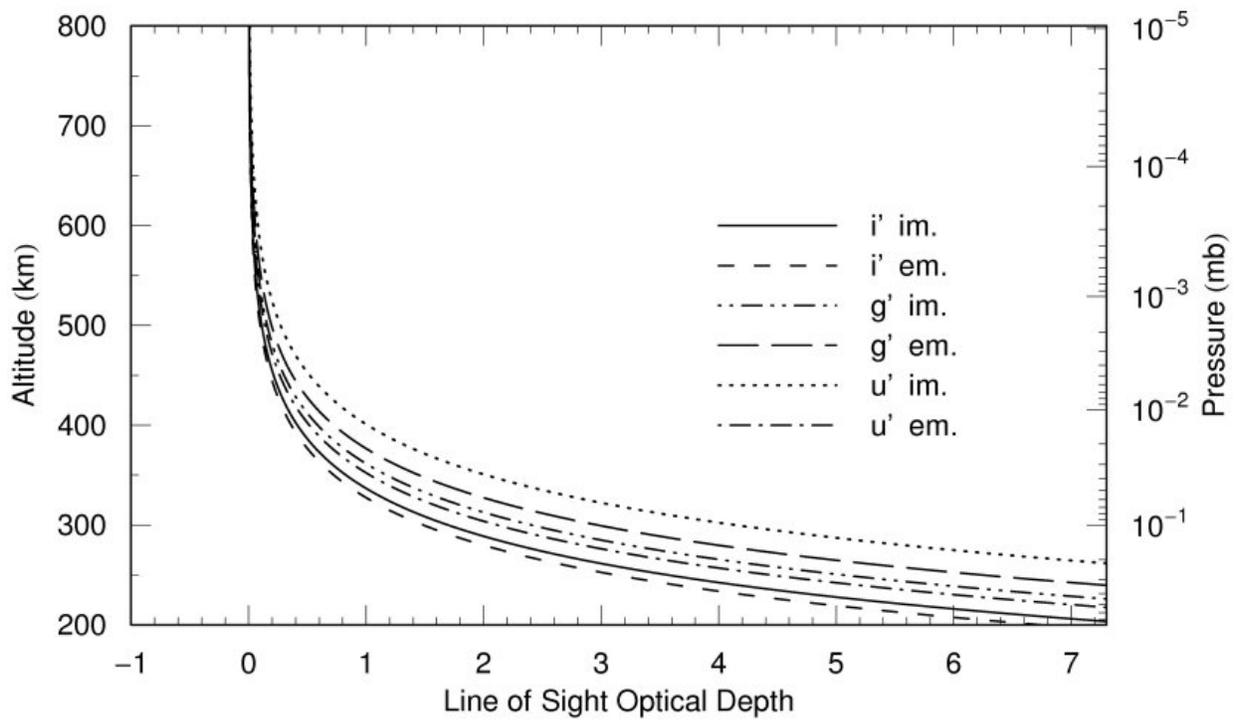

A. Zalucha, Figure 11

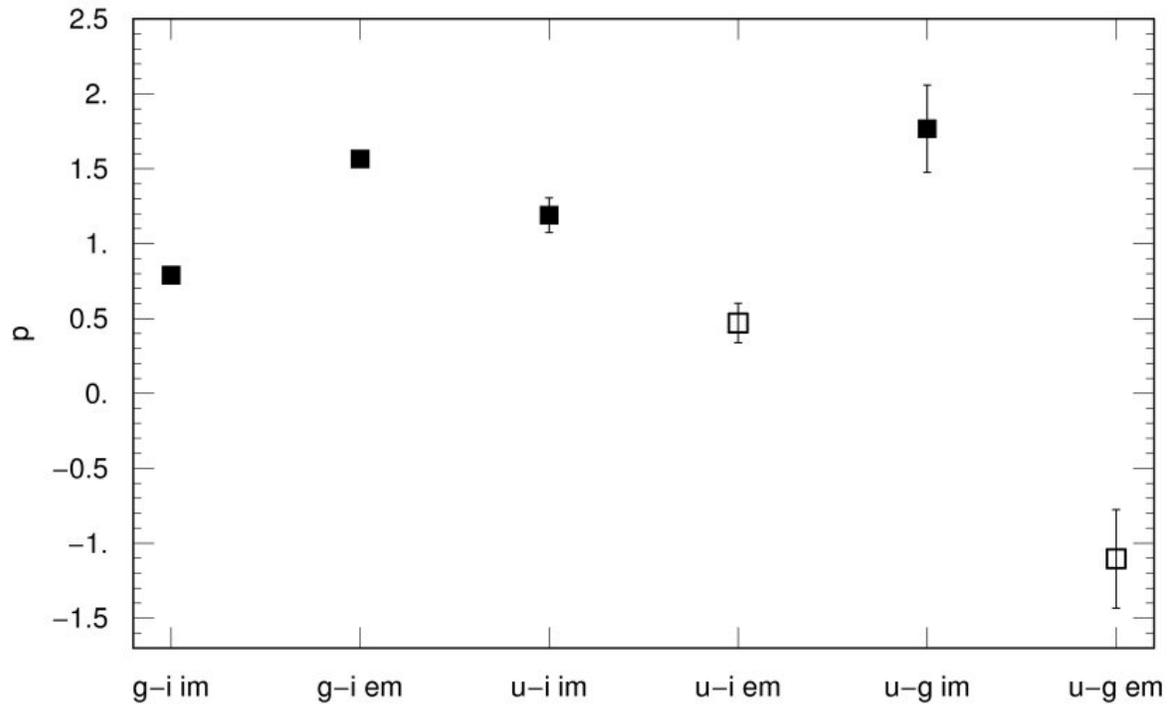

A. Zalucha, Figure 12

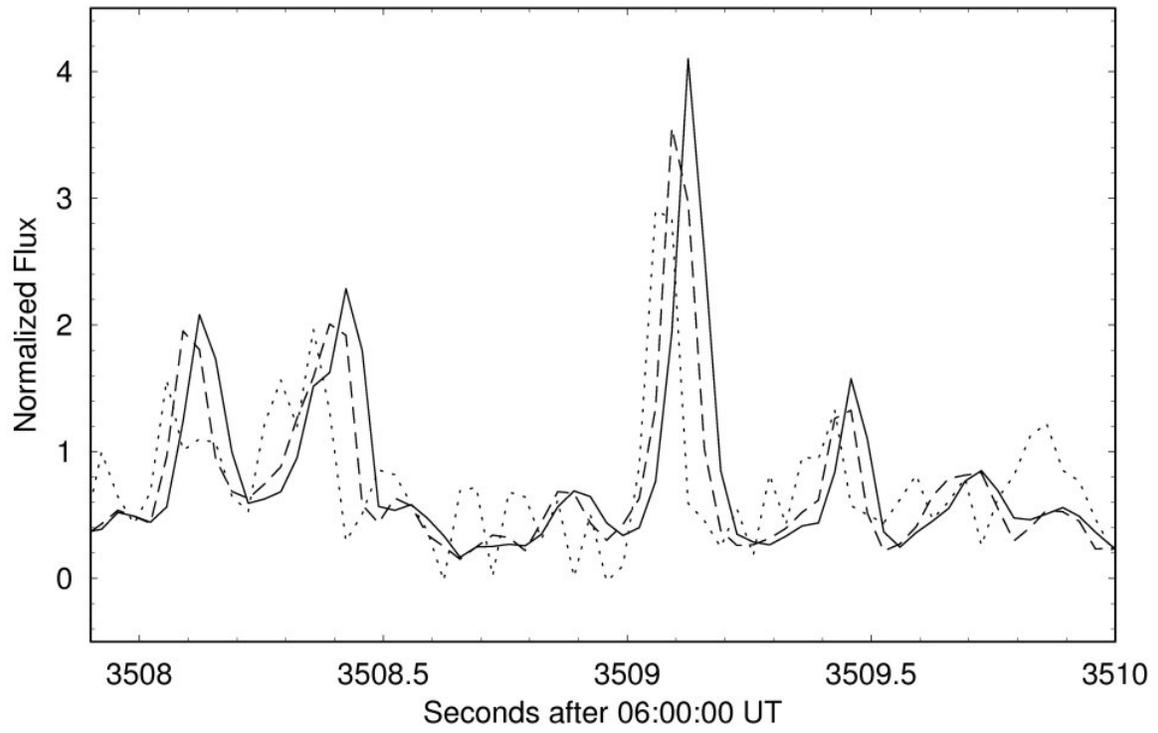

A. Zalucha, Figure 13

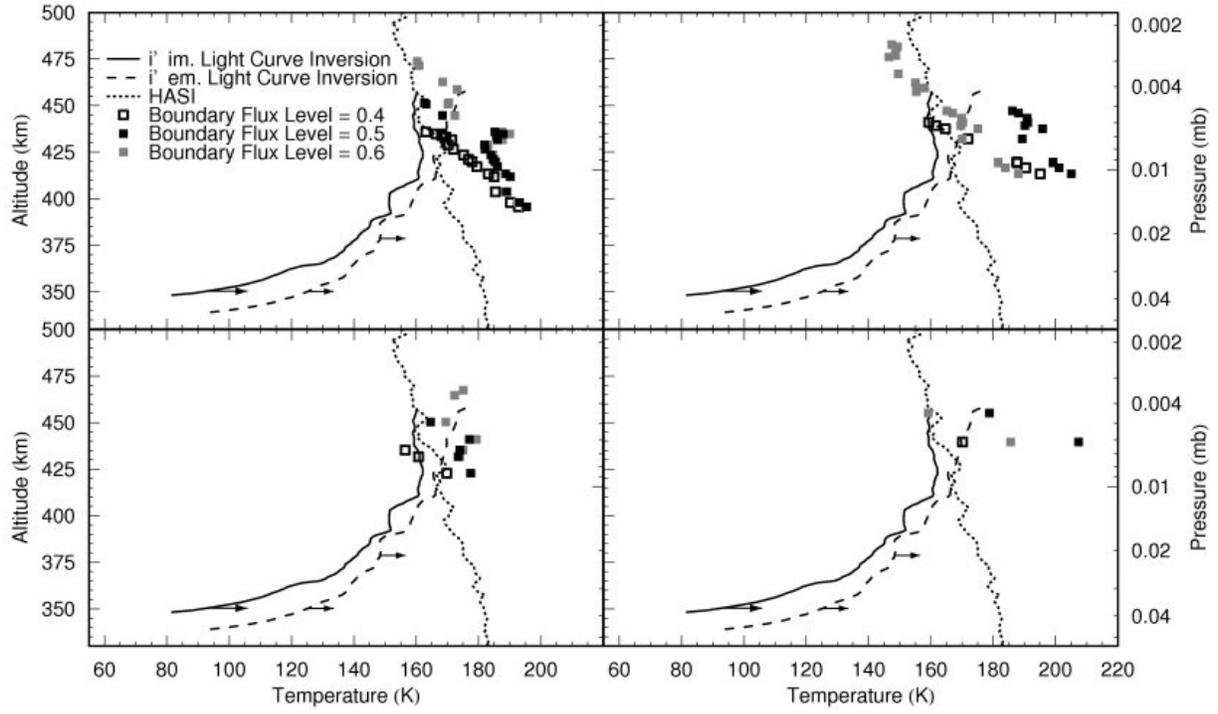

A. Zalucha, Figure 14

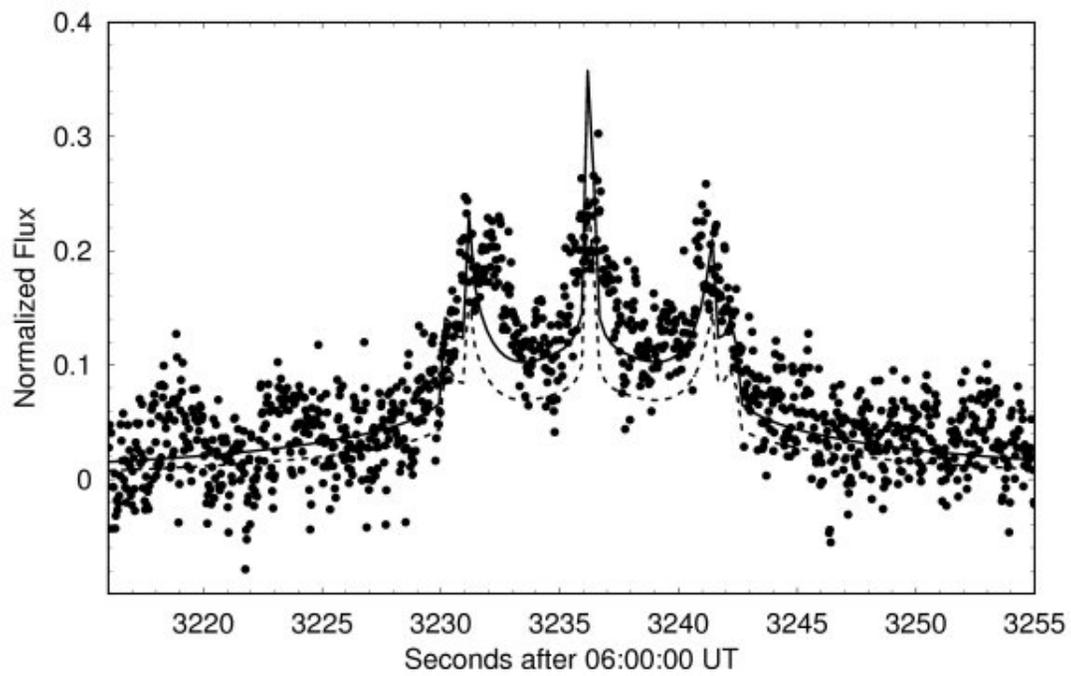